\documentclass[useAMS,usenatbib]{mn2e}

\usepackage{graphicx}
\usepackage{amsmath}
\usepackage{bm}
\usepackage{epsfig}
\usepackage{amssymb}
\usepackage{natbib}
\usepackage{threeparttable} 
\usepackage{subfigure}
\usepackage{setspace}
\usepackage{multirow}    
\usepackage{appendix}
\def\simmore{\mathbin{\lower 3pt\hbox
     {$\rlap{\raise 5pt\hbox{$\char'076$}}\mathchar"7218$}}}   

\title[Spectral analysis of 4U 1728$-$34]{Ionisation state of the accretion disc in the neutron-star low-mass X-ray binary 4U 1728$-$34}
\author[Ming Lyu et al.]
{Ming Lyu$^{1,2}$ \thanks{E-mail: lvming@xtu.edu.cn}, Mariano M\'endez$^2$, Jian-Fu Zhang$^1$ and Fu-Yuan Xiang$^1$ \\
$^1$Department of Physics, Xiangtan University, Xiangtan, Hunan 411105, China\\
$^2$Kapteyn Astronomical Institute, University of Groningen, PO BOX 800, NL-9700 AV Groningen, the Netherlands\\
}

\begin{document}

\date{Accepted XXXX. Received XXXX; in original form XXXX}

\maketitle

\label{firstpage}

\begin{abstract}

We analysed an XMM-Newton plus a simultaneous Rossi X-ray Timing Explorer observation and a separate Suzaku observation of the neutron-star low-mass X-ray binary 4U 1728--34. We fitted the X-ray spectra with the self-consistent reflection model {\sc relxill}. We found that the inclination angle of 4U 1728–34 is 49$\pm$5 degrees, consistent with the upper limit of 60 degrees deduced from the absence of eclipses or dips in this source. The inclination angle in the fit of the XMM-Newton/RXTE observation is larger than 85 degrees, which may be due to the possible calibration issues of the PN instrument in timing mode. We also found that the thermal emission from the accretion disc is not significant. This could be explained either by the relatively high column density of the interstellar medium along the line of sight to the source, which decreases the number of soft disc photons, or if most of the soft thermal photons from the disc are reprocessed in the corona. The ionisation parameter derived from the fits is larger than the value predicted in the framework of the standard reflection model, wherein the disc is irradiated by an X-ray source above the compact object. This inconsistency suggests that irradiation from the neutron star and its boundary layer may play an important role in the ionisation of the accretion disc, and hence in the reflection component in this source.

\end{abstract}


\section{Introduction}  

Low-mass X-ray binary systems (LMXBs) consist of a central compact object (a neutron star or a black hole) accreting matter though an accretion disc from a low-mass (M$\le$1 M$_\odot$) donor star. According to the path they trace in an X-ray colour-colour diagram (CD) or hardness-intensity diagram (HID), \citet{hasinger89} classified the neutron-star LMXBs (NS LMXBs) into two classes, the Atoll and the Z sources. The Z sources ($\sim$0.5-1 $L_{\rm Edd}$) are brighter than the Atoll sources \citep[0.01-0.2 $L_{\rm Edd}$; e.g.][]{ford2000,done07,homan07}. The Atoll sources display three branches in the CD: the island, the lower and the upper banana. The branches in the CD correspond to different spectral states, with mass accretion rate increasing from the island (hard spectral state) to the upper banana branch (soft spectral state). 

In a typical LMXB system, the emission can be decomposed into the contribution of two main components: (i) A soft, thermal, component due to the emission from the accretion disc and, when appropriate, the neutron-star surface and its boundary layer, and (ii) a hard component due to inverse Compton scattering of soft photons from the thermal component(s) by hot electrons in a corona. In the hard spectral state, the inner part of the accretion disc is relatively cool, $\sim$0.3-0.5 keV \citep[e.g.][]{sanna13,lyu14}, and contributes less than 20\% of the total emission between 0.5 and 20 keV. The energy spectrum in the hard state is dominated by a Comptonised component, which is usually described by a power-law-like model with a photon index of $\sim$ $1.6-2.5$ \citep{yoshida93,mariano97}. In the truncated disc scenario \citep[e.g.][and references therein]{done07}, the accretion disc in the hard state is truncated far from the central object, thus leading to a relative low inner-disc temperature and disc flux, while in the soft spectral state, the soft thermal emission becomes dominant. There are two possible sources of thermal emission in a neutron star system, either the surface of the neutron star (plus its boundary layer) or the accretion disc. The disc in the soft state extends down to the inner most stable circular orbit \citep{ss73}, leading to a high inner-disc temperature of $0.7-1.0$ keV \citep{sanna13,lyu14} and a strong thermal component. The electrons in the corona are efficiently cooled down through the inverse Compton scattering process, with seed photons coming from the thermal components. As a consequence, the Comptonised spectrum becomes steep and has a photon index of $\sim$$2-2.5$ \citep{miyamoto93,mariano97}, and little hard emission is detected in some cases \citep{gierliski03,zhang17}.   

Many models have been proposed to explain the NS LMXB spectra. Historically, two main types of models were widely used: (i) the `eastern model' consists of a multi-temperature thermal emission from the accretion disc plus a Comptonised component from an optically thin but geometrically thick corona \citep{mitsuda84,mitsuda89}; (ii) the `western model' consists of a single-temperature blackbody component from the surface of the neutron star (and its boundary layer) and a Comptonised spectrum due to inverse Compton scattering of thermal photons off the hot electrons in the corona \citep{white86}.
  
  4U 1728--34 was first detected by UHURU scans of the Galactic center region in 1976 \citep{forman76}. Later \citet{lewin76} and \citet{hoffman76} detected type I X-ray bursts from the source, identifying the neutron-star nature of the compact object. 4U 1728--34 was classified as an Atoll NS LMXB \citep{hasinger89}, at an estimated distance of $4.4-5.1$ kpc, deduced from measurements of the peak flux of photospheric radius expansion bursts \citep{disalvo00,galloway03}. \citet{migliari03} found a coupling between the disc and the jet in 4U 1728--34 based on a significant correlation between the radio flux density and the X-ray flux. Spectral analysis of 4U 1728--34 has been performed using observations from many satellites in the past, such as Einstein \citep{grindlay81}, SAS--3 \citep{basinska84}, EXOSAT \citep{white86}, SIGMA \citep{claret94}, ROSAT \citep{schulz99}, BeppoSAX \citep{piraino00,disalvo00}, ASCA \citep{narita01}, RXTE, Chandra \citep{dai06}, INTEGRAL \citep{falanga06} and XMM-Newton \citep{ng10,egron11}, BeppoSAX, RXTE \citep{seifina11}, INTEGRAL, RXTE \citep{tarana11}, NuSTAR, Swift \citep{sleator16,mondal17}. 
  
 In this work we used an XMM-Newton observation plus a simultaneous Rossi X-ray Timing Explorer (RXTE) observation, and a Suzaku observation to explore the spectral properties of 4U 1728--34. We applied full reflection models to investigate the spectral features of the source. Furthermore, we studied the ionisation state of the accretion disc by comparing the results from spectroscopy and theory, and we explored the possible mechanisms behind the emission spectrum in this source. The paper is organized as follows: We describe the observations and the details of the data reduction in Section 2. In Section 3 and Section 4, we show the spectral analysis process and the results, respectively. Finally, in Section 5 we discuss our findings.

\section{Observations and data reduction}

In this work we used data from three satellites: An XMM-Newton observation (ObsID:0671180201) plus a simultaneous RXTE observation (ObsID: 96322-01-01-00) taken on 2011-08-28, and a Suzaku observation (ObsID:405048010) obtained on 2010-10-04. We analyzed the XMM-Newton/RXTE and the Suzaku observation separately in this work.

The XMM-Newton observation was taken with the European Photon Imaging Camera, EPIC-PN \citep{xmm01} in timing mode, with a total exposure time of about 52 ks. We used the Science Analysis System (SAS) version 16.1.0 to reduce the PN data with the latest calibration files. We applied the tool {\tt epproc} to extract calibrated events, and converted the arrival time of photons to the barycenter of the solar system using the command {\tt barycen}. We excluded all events at the edge of a CCD and close to a bad pixel, and selected only single and double events for the extraction. We estimated the pileup effect using the command {\tt epatplot}, as suggested by the SAS thread. When we exclude the central one, three and five columns, the 0.5-2.0 keV observed-to-model fraction for doubles is 1.050$\pm$0.003, 1.022$\pm$0.004, and 0.990$\pm$0.004, respectively. We finally selected events within a  41-column wide region centered on the source position, excluding the central five columns to eliminate pile up. The EPIC pn spectrum of the full region has a count rate of $\sim$ 247 cts/s, whereas the one excluding the central five columns has a count rate of $\sim$ 70 cts/s. We removed all X-ray bursts before we extracted the spectra. For PN timing mode observation, the point spread function (PSF) of the instrument extends further than the CCD boundaries, thus the whole CCD was contaminated by the source photons \citep{ng10,hiemstra11}. To model the background, we used the observation of the neutron-star LMXB 4U 1608--52 (ObsID 0074140201) in timing mode when the source was close to quiescence. We excluded the bad time interval with flaring particle background and then extracted the background from a region far from the center of the PSF (RAWX in [2:5]). We produced the response matrices and ancillary response files with the commands {\tt rmfgen} and {\tt arfgen}, respectively. Finally, we rebinned the spectrum with the command {\tt specgroup} to ensure a minimum of 25 counts in every bin and a maximum oversampling factor of 3.

For the RXTE observation we used the Proportional Counter Array \citep[PCA;][]{jahoda06} data only,  since the other instrument, the High Energy X-ray Timing Experiment \citep[HEXTE;][]{roth98}, was not in a good working condition after 2009. We reduced the data using the {\sc heasoft} package version 6.16 according to the RXTE cook book\footnote{http://heasarc.gsfc.nasa.gov/docs/xte/recipes/cook\_book.html}. We applied the tool {\tt saextrct} to extract PCA spectra from Standard-2 data, where only the events from the best calibrated detector, PCU2, were selected. We ran the commands {\tt pcabackest} to generate the PCA background files and {\tt pcarsp} to produce the response files. Finally, we applied the dead time correction to the spectra.

We used the Suzaku data taken with two instruments onboard: the X-ray Imaging Spectrometer (XIS) and the Hard X-ray Detector (HXD). The XIS and HXD detectors cover an energy range of $0.2-12$ keV and $10-70$ keV, respectively. The XIS data were collected by two front-illuminated (FI) detectors (XIS0 and XIS3) and one back-illuminated (BI) detector (XIS1). The 1/4 window option and the burst option were applied to the XIS detectors to limit possible pileup effects.  

We followed exactly the steps from the Suzaku Data Reduction Guide\footnote{http://heasarc.gsfc.nasa.gov/docs/suzaku/analysis/abc/} to reduce all Suzaku data. We used the tool {\tt aepipeline} to recalibrate the XIS and HXD events with the latest calibration files, removed bad pixels and applied the standard GTI selection. After excluding X-ray bursts, we ran the {\sc heasoft} tool {\tt xselect} to extract XIS spectra from a circular region centered at the position of the source. We found that there was no pile up in the spectra. The redistribution matrix files (RMFs) and ancillary response files (ARFs) were generated using the tool {\tt xisrmfgen} and {\tt xissimarfgen}, respectively. Finally, we used the command {\tt addascaspec} to combine the spectra of XIS1 and XIS3 to produce the FI spectra. For the HXD-PIN data, we applied the tool {\tt hxdpinxbpi} to produce the source and background spectra. We applied the dead-time correction to the source spectrum using the pseudo-events files. Since the non X-ray background (NXB) has a count rate 10 times higher than the real background, in order to reduce the Poisson noise, we adjusted the exposure time of the NXB spectra by a factor of 10. Furthermore, a cosmic X-ray background (CXB) was simulated and added to the NXB to generate a total background spectrum. Finally, we downloaded the response file from the online CALDB according to the Suzaku Data Reduction Guide.

\section{Spectral Analysis}
 In this work we used XSPEC version 12.10.0 \citep{arnaud96} to fit the PN and PCA spectra together in the $0.9-25$ keV energy range (PN: $0.9-10$ keV; PCA: $10-25$ keV), and the Suzaku spectra in the 1-50 keV energy range (FI/BI: $1-10$ keV; HXD-PIN: $10-50$ keV). We used the component {\sc phabs} to describe the interstellar absorption along the line of sight, with the solar abundance table of \citet{wilms00} and the photoionisation cross section table of \citet{verner96}. A multiplicative factor was applied to the model to account for possible calibration difference between different instruments. We fixed this factor to 1 for the PN and FI spectrum, and let it free for the spectrum of the other instruments. Finally, we added a 0.5\% systematic error to all the spectra to account for possible effects of the cross calibration on the spectral shape \citep[e.g.][]{sanna13,dai15}. Below we describe the models that we fitted to the spectra, and we present the corresponding fitting results in the next section.
 
\subsection{Direct emission}
We first tried a thermal component plus a Comptonised component to fit the spectra \citep[e.g.][]{sleator16,mondal17}. We selected the model {\sc bbody} to describe the thermal emission from the neutron-star surface and its boundary layer. For the Comptonised component we used the thermal comptonised component {\sc nthcomp} \citep{zdzi96,zyck99}, which describes more accurately the high energy shape and the low energy rollover than an exponentially cutoff power-law component. In order to test whether this combination is able to fit the continuum well, we excluded the 5 keV to 8 keV energy range where significant residuals caused by iron emission line were present (see $\S 3.2$ below). We found that the continuum could be well fitted by the {\sc bbody+nthcomp} model, with a reduced chi-squared value/number of degrees of freedom of 0.99/128 and 1.26/2592 for the XMM-Newton/RXTE observation and the Suzaku observation, respectively. We also tried to add the component {\sc diskbb} \citep{mitsuda84,maki86} to fit the possible emission from an accretion disc, and linked the temperature of the accretion disc, $kT_{dbb}$, in the {\sc diskbb} component to the seed photon temperature, $kT_{seed}$, in the {\sc nthcomp} component. We found that the additional component {\sc diskbb} did not significantly improve the fits: The reduced chi-squared/number of degrees of freedom for the fits were 0.99/127 and 1.21/2591 for the XMM-Newton/RXTE and the Suzaku observation, respectively. The {\sc diskbb} normalization $N_{dbb}$, in the Suzaku observation in this case became extremely large (208132$_{-65410}^{+99428}$), which is likely caused by a degeneracy of the parameters in the model. Therefore, we did not use the {\sc diskbb} component in the rest fitting process \citep[see also,][]{piraino00,dai06,seifina11}.

The seed photons for the thermal Compton process in the corona could either come from the accretion disc or the neutron-star surface and its boundary layer. \citet{sanna13} explored the origin of the seed photons by linking the seed photon temperature, $kT_{seed}$, in the {\sc nthcomp} to either the temperature of the accretion disc, $kT_{dbb}$, in the {\sc diskbb}, or the temperature of the neutron-star, $kT_{bb}$, in the {\sc bbody}, respectively. They found that both options gave statistically acceptable fits, however the blackbody emission became negligible when linking $kT_{bb}$ in the {\sc bbody} to $kT_{seed}$ in the {\sc nthcomp}. We also tried to link $kT_{seed}$ in the {\sc nthcomp} component to the temperature of the neutron-star, $kT_{bb}$,  however, in this case the fitting became worse and $kT_{bb}$ decreased to $\sim$ 0.33 keV or pegged at the lower boundary 0.3 keV for the XMM-Newton/RXTE and the Suzaku observation, respectively. We therefore assumed that the thermal photons from the accretion disc are the seed photons for the Compton process in the corona \citep[e.g.][]{sanna13,lyu14}, and we therefore left the parameter $kT_{seed}$ free to vary, because, as we explained above, we did not detected the {\sc diskbb} component.

\subsection{Phenomenological reflection model of the line emission}
After fitting the continuum we found prominent residuals in the iron line region. The existence of residuals around $5-8$ keV in the fits indicates that reflection from the accretion disc may be present, therefore we added a {\sc gaussian} component to fit the possible iron line emission, constraining the energy of the line to be in the 6.4 $-$ 6.97 keV range. We found that the derived {\sc gaussian} line profiles were in general broad ($\sigma\sim$0.8 keV), so we then substituted the {\sc gaussian} component with a relativistically broadened {\sc laor} line \citep{laor91} 
model, which was developed to describe the emission from a line emitted from an accretion disc around a maximally-rotating black hole. The parameters of the {\sc laor} component are the emissivity index of the disc, $\beta$, the inner and outer radii of the disc, $R_{in}$ and $R_{out}$, respectively, the central energy of the emission line, $E_{\rm line}$, and the inclination angle of the disc with respect to the line of sight, $i$. We found that the best-fitting parameters did not change significantly when the outer disc radius was fixed at 400 $R_{g}$ compared with the ones when it was free to change. We then fixed the outer disc radius in the model at 400 $R_{g}$, where $R_{g}$ = $GM/c^{2}$, with $G$ being Newton’s constant, $c$ the speed of light, and $M$ the mass of the neutron star.

\begin{figure*}
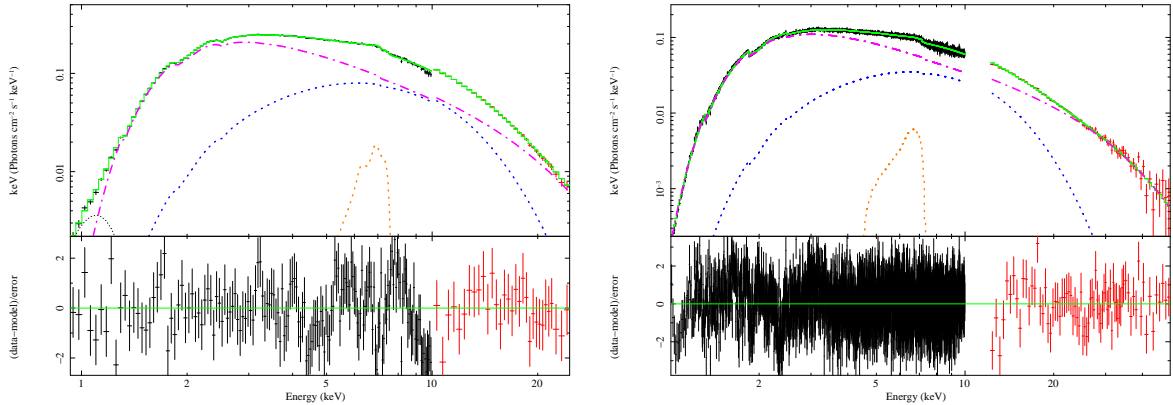

\center
\includegraphics[width=0.3\textwidth,angle=-90]{Fig1_1.eps}
\includegraphics[width=0.3\textwidth,angle=-90]{Fig1_2.eps}
\caption{Fitting results with the phenomenological reflection model {\sc phabs*(bbody+nthcomp+laor)} for the XMM-Newton/RXTE (left) and the Suzaku (right) observations of 4U 1728–34. Each plot shows the fitted spectra and the individual model components (main panel), and the residuals in terms of sigmas (sub panel). The component {\sc bbody}, {\sc nthcomp}, {\sc laor} and the sum of all these model components are plotted with a blue dotted, magenta dashed-dotted, yellow dotted line and green line, respectively. Since the FI and BI spectra in the Suzaku observation are mostly on top of each other in the plot, here we do not plot the BI spectra for clarity. The residuals in the plots are rebinned for plotting purposes.}
\label{ironline}
\end{figure*}

\subsection{Full reflection models}

When X-rays illuminate the accretion disc, the surface of the accretion disc is expected to be ionised, producing a reflection spectrum including fluorescence lines, recombination and other emissions \citep[e.g.][]{ross05}. The shape of the reflection spectrum is influenced by the ionisation state of the disc, and thus the reflection spectrum is important for understanding the thermal balance of the disc. Therefore, we also fitted the broad-band spectrum with a self-consistent reflection model. We first applied the reflection model {\sc relxill} \citep{garcia14,dauser16a}, which describes the reflection off the disc illuminated by a power-law source. The {\sc relxill} model combines the relativistic convolution kernel {\sc relconv} \citep{dauser10} with the reflection grid {\sc xillver} \citep{garcia13}, and it calculates the reflection spectrum for each emission angle. The fit parameters of the {\sc relxill} model are the inclination of the accretion disc, $i$, the dimensionless spin parameter, $a$, the redshift to the source, $z$, the inner and outer radii of the disc, $R_{in}$ and $R_{out}$, respectively, the inner and outer emissivity index of the accretion disc, $q_{in}$ and $q_{out}$, the breaking radius $R_{br}$ where the emissivity changes, the ionisation parameter, $\xi$, the iron abundance, $A_{Fe}$, the photon index of the power-law, $\Gamma$, the cut-off energy of the power-law, $E_{cut}$, the reflection fraction, $R_{refl}$, and the normalization. We fixed $a$ to 0.17 calculated as $a$ = 0.47/$P_{ms}$ \citep{braje00}, where $P_{ms}$=1000/363 ms \citep{stroh96} is the spin period of the neutron star in milliseconds. The inner radius, $R_{in}$, was set to be larger than 5.44 $R_{g}$, while $R_{out}$ and $R_{br}$ were fixed at 400 $R_{g}$, and hence $q_{in}$ and $q_{out}$ were linked to vary together. The redshift $z$ was set to zero and $A_{Fe}$ was fixed to the solar abundance. 

In order to explore the possible geometry of 4U 1728$-$34, we also fitted the spectra with another reflection model, {\sc relxilllp} \citep{garcia14,dauser16}. The {\sc relxilllp} model is a relativistic reflection model for the lamppost geometry, which assumes that the corona  is located above the accretion disc along the spin axis of the compact object. The {\sc relxilllp} model has the same parameters as {\sc relxill}, but replaces the emissivity indeces and the breaking radius with a new parameter, $h$, which is the height of the corona. We set the parameter $fixReflFrac=0$ to fit the reflection freely, and set the rest of the parameters in this model to the same values as in {\sc relxill}.

To sum up, we used the model {\sc phabs$\ast$(bbody+laor+\\nthcomp)} to describe the continuum and iron emission line, and the models, {\sc phabs$\ast$(bbody+relxill)} and {\sc phabs$\ast$\\(bbody+relxilllp)}, to model the full reflection spectra. We found that the inclination angle in the XMM-Newton/RXTE observation could not be well constrained (it was larger than 85 degress) in the fits, and hence we fixed it at 50 degrees, derived from the fits of the full reflection models to the Suzaku observation.

\section{Results}

\begin{table*}
\centering
\caption{Best-fitting results for the fit to the X-ray spectra of 4U 1728--34 with the phenomenological model. The inclination angle in the XMM-Newton/RXTE observation could not be well constrained, so we fixed it to the value in the Suzaku observation when we fit it with the full reflection model (see text). Here we give the unabsorbed flux (erg cm$^{-2}$ s$^{-1}$) in the energy range $0.1-100$ keV. All errors in the Tables are at the 90\% confidence level unless otherwise indicated. A symbol $^{*}$ means that the error pegged at the hard limit of the parameter range.}
\begin{tabular}{|c|c|c|c|}

\hline
Model Comp   &   Parameter      & Suzaku   & XMM-Newton/RXTE   \\
\hline
{\sc phabs}       &$N_{\rm H}$ (10$^{22}$cm$^{-2}$)    &     4.84 $\pm $ 0.06           &      4.41$\pm$ 0.21              \\
{\sc bbody}       &$kT_{\rm BB}$ (keV)                 &     2.2 $\pm $ 0.02            &      2.04$\pm$ 0.03              \\
                  &Norm       (10$^{-3}$)             &     7.1 $\pm $ 0.2             &       15.1$\pm$ 1.1              \\
                  &Flux (10$^{-10}$ c.g.s)                  &       6.1$\pm $ 0.2      &       13.1$\pm$ 0.9                                                     \\
{\sc nthcomp}     &$\Gamma$                           &     2.23 $\pm $ 0.02           &       2.26$\pm$ 0.06             \\
                  &$kT_{\rm e}$ (keV)                  &     7.0 $\pm $ 0.3             &      4.5 $\pm$ 0.4               \\
                  &$kT_{\rm bb}$ (keV)                 &     0.21 $\pm $ 0.04           &      0.32$_{-0.22^{*}}^{+0.08}$      \\
                  &Norm                               &     0.65 $\pm $ 0.03           &       1.02$_{-0.17}^{+0.45}$     \\
                  &Flux (10$^{-10}$ c.g.s)                  &    38.7$\pm $ 2.0        &       60.6$_{-5.5}^{+21.1}$                                               \\
{\sc laor}        &$E_{\rm line}$ (keV)                &     6.97$_{-0.02}^{+0^{*}}$    &      6.59$\pm$ 0.07              \\
                  &$\beta$                            &     3.8 $\pm $ 0.4             &       2.46$\pm$ 0.21             \\
                  &$R_{\rm in}$ ($R_{\rm g}$)          &     6.2 $\pm $ 0.7             &      9.55$\pm$ 3.89              \\
                  &incl $ (^\circ)$                    &     24.7 $\pm $ 1.6            &      50 (fixed)                   \\
                  &Norm (10$^{-3}$)                   &     1.5 $\pm $ 0.2             &       3.78$\pm$ 0.35             \\
                  &Flux (10$^{-10}$ c.g.s)                  &       0.15$\pm $ 0.1     &       0.39$\pm$ 0.03                               \\

\hline
               &$\chi^2_\nu$ ($\chi^2/dof)$     &   1.16 (4758/4093)         &   1.29 (223/172)        \\              
               &Total flux (10$^{-10}$ c.g.s)        &     44.2$_{-1.6}^{+3}$     &    77.3$_{-8.6}^{+18.3}$        \\

\hline          
\end{tabular}   
\medskip        
\\                  
\label{line}     
\end{table*}

\begin{figure*}
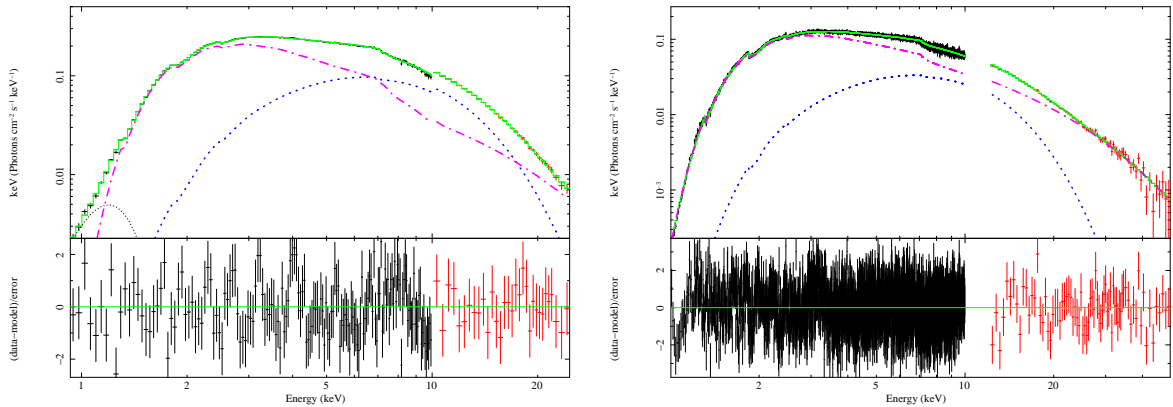

\center
\includegraphics[width=0.3\textwidth,angle=-90]{Fig2_1.eps}
\includegraphics[width=0.3\textwidth,angle=-90]{Fig2_2.eps}
\caption{Fitting results with the full reflection model {\sc phabs*(bbody+relxilllp)} for the XMM-Newton/RXTE (left) and the Suzaku (right) observations of 4U 1728–34. Each plot shows the fitted spectra and the individual model components (main panel), and the residuals in terms of sigmas (sub panel). The component {\sc bbody}, {\sc relxilllp} and the sum of the model components are plotted with a blue dotted, magenta dashed-dotted and green line, respectively. Since the FI and BI spectra in the Suzaku observation are mostly on top of each other in the plot, here we do not plot the BI spectra for clarity. The residuals in the plots are rebinned for plotting purposes.}
\label{lp}
\end{figure*}

In Table \ref{line} we show the fitting results using the {\sc laor} model. The blackbody temperature, $kT_{\rm BB}$, is  2.2$\pm 0.02$ keV in the Suzaku observation and 2.04$\pm 0.03$ keV in the XMM-Newton/RXTE observation. The power-law index, $\Gamma$, is 2.23$\pm 0.02$ and 2.26$\pm 0.06$, while the electron temperature, $kT_{\rm e}$, is 7.0$\pm 0.3$ keV and 4.5 $\pm$ 0.4 keV in the Suzaku and the XMM-Newton/RXTE observation, respectively. The corresponding energy of the {\sc laor} line is 6.97$_{-0.02}^{+0}$ keV and 6.59$\pm 0.07$ keV in the Suzaku and XMM-Newton/RXTE observation, respectively, suggesting that the accretion disc is highly or moderately ionised in these two observations. The corresponding spectra, individual components and residuals of the fits are shown in Figure \ref{ironline}.

In Table \ref{reft} we summarise the fitting results with the reflection model {\sc relxill}. The reduced chi-square with this model is 1.14 and 1.0 for 4094 and 173 degrees of freedom for the Suzaku and XMM-Newton/RXTE observation, respectively. The inclination angle was well constrained at 49$_{-3}^{+8}$ degrees in the Suzaku observation, consistent with the fact that no eclipse has ever been observed in this source. The reflection fraction, $R_{refl}$, is 0.54$\pm$0.08 in the Suzaku observation and 1.39$_{-0.32}^{+0.6}$ in the XMM-Newton/RXTE observation. This may suggest that more photons from the corona illuminate the accretion disc in the XMM-Newton/RXTE than in the Suzaku observation, which is also consistent with the higher ionisation state of the disc in the XMM-Newton/RXTE observation. The ionisation parameter, $log(\xi)$, is 2.71$\pm$0.07 in the Suzaku observation, and 3.92$\pm$0.16 in the XMM-Newton/RXTE observation. The inner radius of the disc, $R_{in}$, is 14.1$_{-2.9}^{+10.7} R_{g}$ and 5.44$_{-0}^{+2.32} R_{g}$ in the Suzaku and XMM-Newton/RXTE observation, respectively.

We show the fitting results with the reflection model {\sc relxilllp} in Table \ref{lpt}. Most of the parameters in the fits with the model {\sc relxilllp} are similar to the ones in the fit with {\sc relxill}, except that the reflection fraction, $R_{refl}$, is systematically higher in  {\sc relxilllp} than in {\sc relxill}. In the case of {\sc relxilllp} $R_{refl}$ is 1.68$_{-0.33}^{+1.19}$ and 3.05$_{-0.65}^{+1.11}$ in the Suzaku and XMM-Newton/RXTE observation, respectively. There is no clear difference in the height of the corona, $h$, in the two observations: $h$ is 15.5$_{-9.2}^{+11.8}$ $R_{g}$ (90\% confidence level) in the Suzaku observation, while it is 22.3$\pm$6.7 $R_{g}$ in the XMM-Newton/RXTE observation. The corresponding spectra, individual components and residuals of the fits are shown in Figure \ref{lp}.

\begin{table*}
\centering
\caption{Best-fitting results for the fit to the X-ray spectra of 4U 1728--34 with the reflection model {\sc phabs*(bbody+relxill)}. We fixed the inclination angle in the XMM-Newton/RXTE observation to the value in the Suzaku observation since this parameter in the XMM-Newton/RXTE observation could not be well constrained. Here we give the unabsorbed flux (erg cm$^{-2}$ s$^{-1}$) in the energy range $0.1-100$ keV. A symbol $^{*}$ means that the error pegged at the hard limit of the parameter range.}
\begin{tabular}{|c|c|c|c|}
\hline
\hline
Model Comp   &   Parameter     & Suzaku   & XMM-Newton/RXTE    \\
\hline

{\sc phabs}    &$N_{\rm H}$ (10$^{22}$cm$^{-2}$)       &  4.92 $\pm $ 0.03          &      5.18  $\pm$ 0.14         \\
{\sc bbody}    &$kT_{\rm BB}$ (keV)                    &  2.27 $\pm $ 0.04          &      2.15  $\pm$ 0.03         \\
               &Norm       (10$^{-3}$)                &  6.9 $\pm $ 0.2            &       19.4  $\pm$ 0.5          \\
               &Flux (10$^{-10}$ c.g.s)                     &    5.4 $\pm $ 0.1    &       16.2$_{-0.2}^{+0.7}$                                                    \\
{\sc relxill}  &$\beta$                               &  2.8$_{-0.2}^{+0.5}$       &       2.32  $\pm$ 0.16          \\
               &incl $(^\circ)$                       &  49$_{-3}^{+8}$        &           50 (fixed)                            \\
               &$R_{\rm in}$ ($R_{\rm g}$)             &  14.1$_{-2.9}^{+10.7}$     &      5.44  $_{-0^{*}}^{+2.32}$      \\
               &$\Gamma$                              &  2.03 $\pm $ 0.04          &       2.43  $\pm$ 0.09        \\
               &log($\xi$)                            &  2.71 $\pm $ 0.07          &       3.92  $\pm$ 0.16        \\
               &E$_{cut}$ (keV)                       &  16.6 $\pm $ 1.2           &       19.95 $_{-2.0}^{+7.03}$ \\
               &R$_{refl}$                            &  0.54 $\pm $ 0.08          &       1.39  $_{-0.32}^{+0.6}$ \\
               &Norm (10$^{-3}$)                      &  5.5  $\pm $ 0.3           &       9.3   $\pm$ 2.4         \\
               &Flux (10$^{-10}$ c.g.s)               &    66.7$_{-1.5}^{+2.8}$   &        206$_{-19}^{+65}$                                    \\

\hline
               &$\chi^2_\nu$($\chi^2/dof)$     &   1.14 (4668/4094)     &  1.0 (173/173)     \\
               &Total flux (10$^{-10}$ c.g.s)        &  73.6$_{-3}^{+1.5}$    &  222$_{-19}^{+66}$     \\
                                                                                                                
\hline
\end{tabular}
\medskip  
\\
\label{reft}
\end{table*}

\begin{table*}
\centering
\caption{Best-fitting results for the fit to the X-ray spectra of 4U 1728--34 with the reflection model {\sc phabs*(bbody+relxilllp)}. We fixed the inclination angle in the XMM-Newton/RXTE observation to the value in the Suzaku observation since this parameter in the XMM-Newton/RXTE observation could not be well constrained. Here we give the unabsorbed flux (erg cm$^{-2}$ s$^{-1}$) in the energy range $0.1-100$ keV. A symbol $^{*}$ means that the error pegged at the hard limit of the parameter range.}
\begin{tabular}{|c|c|c|c|}
\hline
\hline
Model Comp   &   Parameter     & Suzaku   & XMM-Newton/RXTE     \\
\hline

{\sc phabs}      &$N_{\rm H}$(10$^{22}$cm$^{-2}$)  &  4.91 $\pm $ 0.03          &      5.17 $\pm$ 0.12                                   \\
{\sc bbody}      &$kT_{\rm BB}$(keV)               &  2.27 $\pm $ 0.03          &      2.15 $\pm$ 0.02                                   \\
                 &Norm       (10$^{-3}$)           &  6.9 $\pm $ 0.2            &      19.4 $\pm$ 0.4                                  \\
                 &Flux (10$^{-10}$ c.g.s)          &   6.0 $\pm $0.2            &      16.2$_{-0.4}^{+0.7}$                                                                      \\
{\sc relxilllp}  &h ($R_{\rm g}$)               &  15.5$_{-9.2}^{+11.8}$        &      22.3 $\pm$ 6.7                                    \\
                 &incl $(^\circ)$                   &  49$_{-3}^{+6}$           &      50 (fixed)                                                  \\
                 &$R_{\rm in}$($R_{\rm g}$)        &  13.8$_{-3.0}^{+8.7}$      &      5.44 $_{-0^{*}}^{+3.87}$                                  \\
                 &$\Gamma$                         &  2.04 $\pm $ 0.04          &      2.43 $\pm$ 0.10                                   \\
                 &log($\xi$)                       &  2.7 $\pm $ 0.06           &      3.91 $\pm$ 0.15                                   \\
                 &E$_{cut}$ (keV)                  &  16.6 $\pm $ 0.8           &      19.2 $_{-1.8}^{6.3}$                              \\
                 &R$_{refl}$                       &  1.68$_{-0.33}^{+1.19}$    &      3.05 $_{-0.65}^{+1.11}$                           \\
                 &Norm (10$^{-3}$)                 &  7.2$_{-1.2}^{+3.8}$       &      11.8 $\pm$ 3.0                                  \\
                 &Flux (10$^{-10}$ c.g.s)          &   64.6$_{-3.8}^{+2.8}$     &      204$_{-25}^{+43}$         \\
  
\hline
               &$\chi^2_\nu$($\chi^2/dof)$     &   1.14 (4669/4094)     &      1.04 (180/173)              \\
               &Total flux (10$^{-10}$ c.g.s)        & 70.8$\pm $  3.1  &       220$_{-17}^{+41}$                     \\
                                                                                                                
\hline
\end{tabular}
\medskip  
\label{lpt}
\end{table*}

\begin{figure*}
\centering
\includegraphics[width=0.6\textwidth]{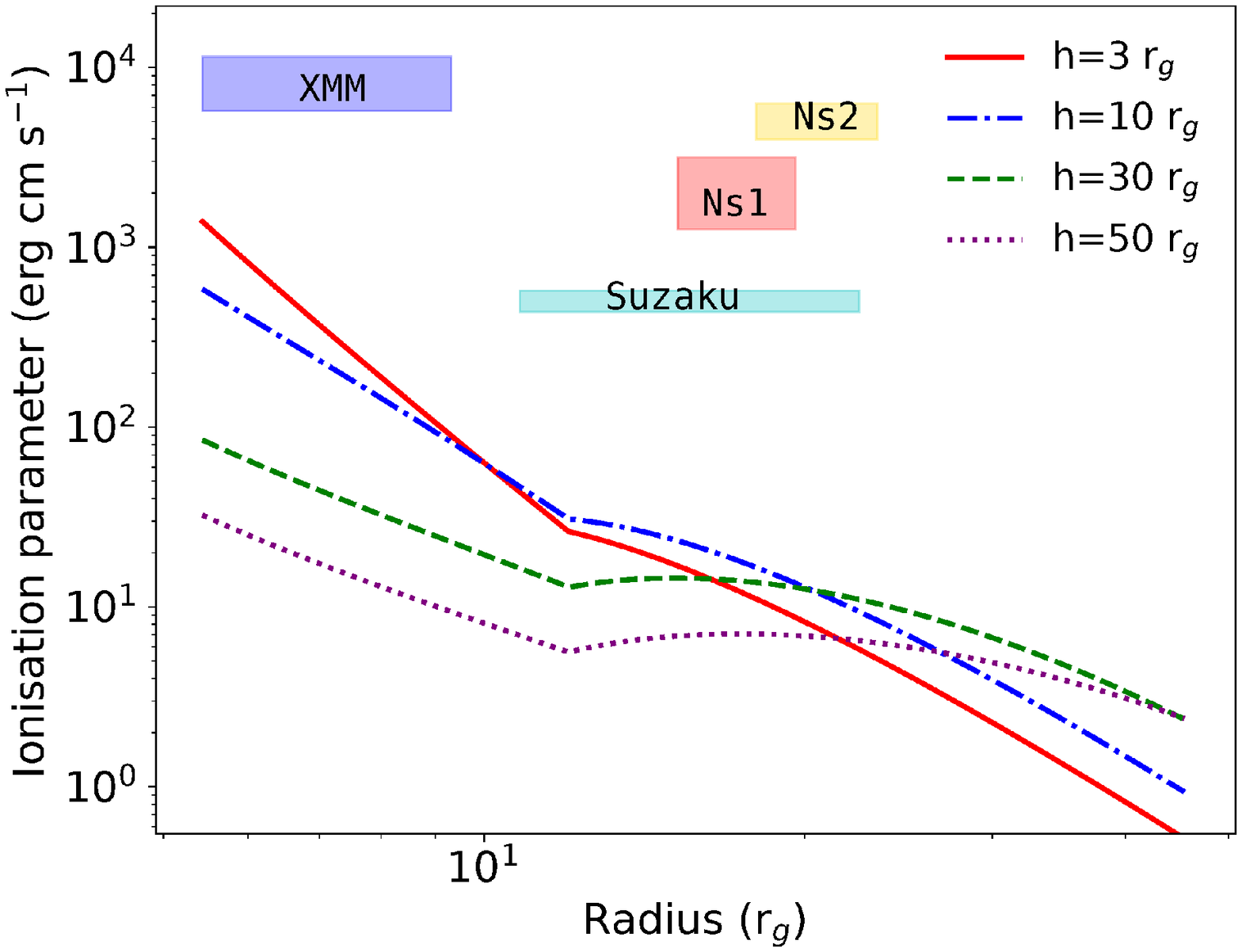}
\caption{The predicted ionisation profiles of an accretion disc illuminated by a lamppost X-ray source located at different heights above the central source. The curves are calculated using eq.(1) in the text, taken from \citet{ballantyne17}, assuming $\eta$=0.1, $\alpha$=0.3, $\lambda$=0.2 and $f$=0.45. The spin parameter is fixed at 0.17 for 4U 1728--34 and $r_{\mathrm{in}}$ and $r_{\mathrm{out}}$ are 5.44 $R_{g}$ and 400 $R_{g}$, respectively. The $R_R, R_z$ and $R_T$ factors at $r< 12$ $R_{g}$ are fixed at their values at $r= 12$ $R_{g}$ to avoid an unphysical break in $\xi(r,h)$. As shown in the legend, ionisation profiles at different heights are marked in different line styles and colours. Here we compared the predicted ionisation profiles with the ionisation derived in this work and in \citet{mondal17}. The two observations in this work were labeled as XMM-Newton and Suzaku, and the two simultaneous NuSTAR and Swift observations in \citet{mondal17} were labeled as Ns1 and Ns2, respectively.}
\label{res}
\end{figure*}

\section{discussion}  

\citet{egron11} analyzed one XMM-Newton observation of 4U 1728--34 with a reflection model and constrained the inclination angle of the system to be within 44 degrees to 60 degress. \citet{mondal17} analyzed two simultaneous NuSTAR and Swift observations of 4U 1728--34 and constrained the inclination angle to be within 22 degrees to 40 degrees. In this work we find that the inclination angle of 4U 1728--34 is about 49$\pm 5$ degrees from the fit to the Suzaku observation, consistent with the range from \citet{egron11} but a bit larger than the value of \citet{mondal17}. We also found that the inclination angle is larger than 85 degrees and could not be well constrained with the XMM-Newton/RXTE data, similar with previous findings \citep[e.g.][]{pandel08,sanna13}. \citet{sanna13} analysed another neutron-star LMXB 4U 1636--53 using six XMM-Newton observations with the PN camera in timing mode, plus simultaneous RXTE observations, and found high inclination angles in the fits, inconsistent with the absence of eclipses and dips in 4U 1636--53. They suggested that the PN timing mode data may be affected by calibration issues, which leads to high inclination angles in the fits. In this work, the derived high inclination angles in the XMM-Newton/RXTE observation may be due to the same calibration issues. When we fit the XMM-Newton spectrum alone, the inclination angle is still larger than 85 degrees. The problem with the calibration of EPIC instrument, would lead to a steeper spectrum, and hence to a higher inclination angle in the fits.

The fitting results show that the {\sc nthcomp} component dominated the broad band energy spectrum in both the XMM-Newton/RXTE and Suzaku observations. This agrees with the conclusion of \citet{seifina11} that the energy spectrum of 4U 1728--34 in all states is dominated by the power-law component. \citet{seifina11} investigated more than 120 RXTE observations of 4U 1728--34, and they found that most of the soft thermal emission from the accretion disc in 4U 1728--34 is reprocessed in the corona, thus the direct disc emission is weak. In this work, the thermal emission from the disc in both the XMM-Newton/RXTE and the Suzaku observation is not required, consistent with their finding. A further calculation shows that the upper limit of the weak {\sc diskbb} component in the fit in this work is able to produce the observed Comptonised component. The absence of the disc emission can also be the consequence of the relatively high column density, $N_{\rm H}$, of the interstellar medium along the line of sight to the source, which reduces the number of soft disc photons that we observe, thus leading to a weak disc component in the observed energy spectra.

\citet{ballantyne17} investigated the ionisation parameter in the accretion disc at a radius $r$ from the black hole, with the disc being irradiated by an X-ray source at a height, $h$, above the black hole symmetry axis. \citet{ballantyne17} developed a formula which takes into account the effects of gravitational light-bending and focusing of radiation onto the disc. According to their calculation, there is a strong ionisation gradient across the surface of the inner disc that depends on the black hole spin and the lamppost height. This model provides a good way to connect the height of the corona with the ionisation state and the inner radius of the accretion disc. For this we applied eq.10 in \citet{ballantyne17}:

\begin{equation}
\begin{aligned}
\xi(r,h)= & (5.44\times 10^{10}) \left ({\eta \over 0.1} \right )^{-2} \left (
  {\alpha \over 0.1} \right) \lambda^3 \left ( {r \over r_g} \right )^{-3/2} R_z^{-2} R_T^{-1} \\
  & \times R_R^3 f(1-f)^3 F(r,h) g_{lp}^2 \mathcal{A}^{-1} \mathrm{erg\ cm\ s^{-1}},
\label{eq:newxi}
\end{aligned}
\end{equation}

where $\xi(r,h)$ is the ionisation parameter of the disc at a radius $r$ from the central source where the illuminating lamppost is at a height $h$, $\eta$ is the radiative efficiency of the accretion process, $\alpha$ is the viscosity parameter, $\lambda$ is the Eddington ratio, $\lambda=L_{bol}/L_{Edd}$, $R_R, R_z$ and $R_T$ are relativistic corrections to the Newtonian $\alpha$-disc equations \citep{krolik99}, $f$ is the coronal dissipation fraction, and $g_{lp}$=$\nu_{\mathrm{disc}}$/$\nu_{\mathrm{src}}$ is the ratio between the measured frequency of a photon striking the disc and its frequency at the source \citep{dauser13}. The function $F(r,h)$ in the equation describes the shape of the irradiation pattern \citep{Fukumura07}, and $\mathcal{A}$ is the integral of $F(r,h)\times g_{\mathrm{lp}}^2$ over the entire disc, $\mathcal{A}=\int_{r_{\mathrm{in}}}^{r_{\mathrm{out}}} F(r,h) g_{\mathrm{lp}}^2 dS(r)$.

For the calculations, we set the $\eta$=0.1, $\alpha$=0.3 \citep{ballantyne17,penna13}, $\lambda$=0.2 and $f$=0.45 \citep{vasud07}. The spin parameter, $a$, is fixed at 0.17 for 4U 1728--34 and $r_{\mathrm{in}}$ and $r_{\mathrm{out}}$ are 5.44 $R_{g}$ and 400 $R_{g}$, respectively. We found that there was a break of the ionisation profiles around $r$=12 $R_{g}$, similar to the one at $r$=4 $R_{g}$ reported in the simulation of a fast-rotating black-hole system in \citet{ballantyne17}. This break is due to the divergence of $R_R$ and $R_T$ as the radius approaches the innermost stable circular orbit (ISCO). We then applied the same procedure as in \citet{ballantyne17} to fix this: the $(R_R, R_z, R_T)$ factors at $r< 12$ $R_{g}$ are fixed at their values at $r= 12$ $R_{g}$.  

Figure \ref{res} shows the ionisation profiles when the illuminating corona is located at different heights above the disc. We compared the ionisation curves predicted by the formula with the values derived from the fits with the model {\sc relxilllp} in this work and the ones from the fits with the model {\sc relxill} to two simultaneous NuSTAR and Swift observations in \citet{mondal17}. As shown in the figure, the ionisation profiles predicted by the model of \citet{ballantyne17} are significantly smaller than the range derived from the observations. The ionisation parameter is below 100 at $r$ $>$ 10 $R_{g}$, while it increases rapidly as the radius further decreases but is still lower than the range from the fits.

The difference between the ionisation parameter predicted by the model and the one deduced from data can be bridged if we changed the value of certain parameters in the model. Notwithstanding, the new values that the parameters must take for the model to qualitatively match the data either contradict some observational result, or are unphysical. For instance, for the model to match the data, the luminosity of the source should be 40\% of the Eddington luminosity, $\lambda = 0.4$ in eq. 1, whereas the spectral fits indicate that the luminosity in this observation was about 20\% Eddington. To make it 40\% Eddington, either the mass of the neutron star needs to be 0.7 M$_\odot$, or the distance should be as large as 7 kpc, which is inconsistent with previous observational results. For instance, \citet{disalvo00} showed that the distance to 4U 1728--34 is about 5.1 kpc assuming a 1.4 M$_\odot$ neutron star. \citet{galloway03} derived a distance of 4.4 (4.8) kpc for a M=1.4 (2.0) M$_\odot$ neutron star with cosmic atmospheric abundance (X=0.7), while the distance could be up to 30\% larger ($\sim$6.2 kpc) if the atmosphere of the neutron star consisted purely of helium. The model would also match the data if we increased the viscosity parameter, $\alpha$, from 0.3 to 1, or decreased the radiative efficiency from 0.1 to 0.03, however, in these cases the adopted values of the parameters deviate significantly from the values deduced in previous work. Typically $\alpha\sim0.1$ \cite[e.g.][]{accretion02}, whereas some work yield even smaller values of this parameter. For example, in the black hole candidate GRS 1915+105, \citet{Belloni97} found that the viscosity parameter is only 0.00004 and 0.0005 for the Schwarzschild and extreme Kerr cases, respectively.

The inconsistency between the ionisation parameter predicted by the model and the ones obtained from the fits may be due to the fact that the disc is also illuminated by radiation from the neutron star and its boundary layer, which is not included in the standard reflection models. \citet{cackett10} investigated the iron emission lines in 10 neutron star LMXBs, and found that illuminating photons likely come from the neutron star and the boundary layer. Their analysis showed that the spectra could be well fitted with a reflection model, with the accretion disc being illuminated by the blackbody component. Besides, they further calculated the maximum height of the boundary layer, $z<24$ km, consistent with the scenario that the boundary layer illuminates a geometrically thin disc. A subsequent analysis of the bright atoll source 4U 1705--44 by \citet{daa10} also showed that the spectrum could be well-fitted by the sum of two thermal components together with a reflection component, wherein the blackbody component provides illuminating photons to the disc. Their analysis provides another example that the reflection component in neutron star LMXBs may come from hard X-ray thermal irradiation, which is likely to be the emission from the boundary layer.

There were several type I X-ray bursts in the observations processed in this work, and hence the burst emission may also illuminate and ionise the accretion disc. \citet{ballantyne04} presented models of X-ray reflection from a constant-density slab illuminated by a blackbody emitted from a neutron star burst. The simulated spectral showed a prominent Fe line, while the reflection profiles drop off quickly above 10 keV compared to the profiles assuming power-law component illumination. Nevertheless, as calculated in \citet{ballantyne04a}, the recombination time for He-like Fe (Fe XXV) is only $\sim$ 10$^{-4}$ s, so the disc will soon return to its previous ionisation state after a burst. Therefore, the bursts will likely have no or little contribution to the average ionisation state of the disc.

\section{Summary} 

In this work we used an XMM-Newton plus simultaneous RXTE observation and a Suzaku observation to study the spectral properties of the neutron-star LMXB 4U 1728--34. We found that the spectra could be well fitted with a model that does not include thermal emission from the accretion disc, The continuum is dominated by the hard component, while both the full reflection model {\sc relxill} and {\sc relxilllp} provide good fits to the spectra. The inclination angle of 4U 1728--34 derived from the Suzaku observation is 49$\pm$5 degrees, while it is not well constrained in the XMM-Newton/RXTE observation. The accretion disc is moderately to highly ionised in these two observations, with the illuminating source located at similar heights in the Suzaku and XMM-Newton/RXTE observation. We found that the ionisation parameter derived in this work and in \citet{mondal17} are larger than, and inconsistent with, the ones predicted by the lamppost model of \citet{ballantyne17}, assuming that the disc is irradiated only by an X-ray source above the compact object. This inconsistency may be due to the effect of the contribution from the neutron star and its boundary layer to the reflected emission, which is not included in the model. A model focusing on the roles of both the the neutron star (its boundary layer) and the corona in the illuminating and ionising process is needed to investigate neutron star LMXBs; in the meantime, high quality data from observations is also required in order to break possible degeneracy problems in the spectral analysis.

This research has made use of data obtained from the High Energy Astrophysics Science Archive Research Center (HEASARC), provided by NASA's Goddard Space Flight Center. This research made use of NASA's Astrophysics Data System. Lyu is supported by National Natural Science Foundation of China (grant No.11803025); and the Hunan Provincial Natural Science Foundation (grant No. 2018JJ3483) and Hunan Education Department Foundation (grant No. 17C1520). F.Y.X. is supported by the Joint Research Funds in Astronomy (U1531108 and U1731106). J.F.Z. thanks the supports from the National Natural Science Foundation of China (grant No.11703020); and the Hunan Provincial Natural Science Foundation (grant No. 2018JJ3484).

\clearpage

\bibliographystyle{mn}
\bibliography{biblio}

\label{lastpage}

\end{document}